\documentclass{article}
\usepackage[pdftex]{graphicx}
\usepackage{natbib}
\usepackage[utf8]{inputenc}
\usepackage{authblk}

\usepackage{multirow}
\usepackage{hyperref}
\usepackage{indentfirst}
\usepackage{amsfonts, amssymb, amsmath, amsthm, bm} 
\usepackage{dsfont, accents}

\usepackage{float, enumerate}
\usepackage[mathscr]{euscript}

\usepackage{caption}
\usepackage{subcaption}
\usepackage[section]{placeins}

\usepackage{booktabs} 

\title{Event-Based Limit Order Book Simulation under a Neural Hawkes Process: Application in Market-Making}
\author[1]{Luca Lalor}
\author[1]{Anatoliy Swishchuk}
\affil[1]{\small University of Calgary, Department of Mathematics and Statistics, Calgary, AB T2N 1N4, Canada}

\date{\today}

\newcommand{\secondlevelheading}[1]{\textbf{\textit{#1}}}

\begin{document}
\maketitle
\thispagestyle{empty}

\begin{abstract}
In this paper, we propose an event-driven Limit Order Book (LOB) model that captures twelve of the most observed LOB events in exchange-based financial markets. To model these events, we propose using the state-of-the-art Neural Hawkes process, a more robust alternative to traditional Hawkes process models. More specifically, this model captures the dynamic relationships between different event types, particularly their long- and short-term interactions, using a Long Short-Term Memory neural network. Using this framework, we construct a midprice process that captures the event-driven behavior of the LOB by simulating high-frequency dynamics like how they appear in real financial markets. The empirical results show that our model captures many of the broader characteristics of the price fluctuations, particularly in terms of their overall volatility. We apply this LOB simulation model within a Deep Reinforcement Learning Market-Making framework, where the trading agent can now complete trade order fills in a manner that closely resembles real-market trade execution. Here, we also compare the results of the simulated model with those from real data, highlighting how the overall performance and the distribution of trade order fills closely align with the same analysis on real data.
\end{abstract}


\pagenumbering{arabic}

\newpage

\section{Introduction}
There are many works devoted to modeling limit order book (LOB) data, which often follow some form of approximation process for the midprice\footnote{The midprice is the midpoint between the best bid and ask price.} process. A detailed overview of LOB models can be found in the survey paper by \cite{gould2013limit}, and also in a more recent study by \cite{jain2024limit}, where much of the current empirical evidence in the popular literature is presented, along with a clear indication that there are still many limitations, not limited to just the model of the price process. In this paper, we construct a midprice process from the ground up, using real LOB events to replicate its evolution in the market. In the LOB system, the three main event types are limit order entries, market order entries that lead to trade order executions, and limit order cancellations. Each time one of these event types occurs, they are given a unique timestamp, almost surely, and they occur at non-uniform discrete time points. We believe that a model that can account for this event-based structure is more in line with the realities of LOB modeling for Algorithmic and High-Frequency Trading (HFT) strategies. 

Popular LOB models in the literature often rely on approximations for pricing, typically ignoring the discrete-time, event-driven nature of real LOB dynamics. This introduces limitations beyond merely misapproximating the midprice process, potentially leading to greater inaccuracies when simulating the performance of algorithmic and HFT strategies. For instance, as shown in \cite{law2019market}, many models that include some form of diffusion-based pricing model (with some popular and regularly cited works including \cite{bertsimas1998optimal}, \cite{bouchard2011optimal}, \cite{cartea2015algorithmic}, \cite{gueant2017optimal}, often lead to so-called ``phantom gains" appearing in Algorithmic and HFT applications, particularly under trading strategies that involve executing numerous limit orders, such as in Market-Making (MM). When backtesting or implementing an MM strategy, it becomes evident that identifying which LOB event(s) caused a price movement is crucial. However, random-walk models like Brownian motion cannot capture this, as LOB events are entirely independent of the modeled diffusion process. One simple example that could help clarify this for the reader would be to think about a MM agent who posts limit orders on the best bid/ask. Their limit orders are much more likely to be executed when price goes against them than in their favor, which a diffusion-based model cannot account for. Thus, simulating a MM strategy within a diffusion model often overestimates the ease of obtaining favorable execution for limit orders compared to real-world conditions. Empirical evidence in line with this particular example can be found in \cite{lalor2024market} and \cite{delise2024negative}, where the adverse selection problem in the LOB that underpins this particular trading strategy is studied. Diffusion-based pricing models are widely used in the mathematical finance, algorithmic trading, and HFT literature due to their simplicity in modeling and simulation. However, their inherent limitations become more pronounced in HFT contexts, reducing their applicability in real-world markets.

In this study, we propose a novel approach that models 12 LOB events using a nonlinear multivariate Hawkes process (MVHP), simulated within a standard Neural Hawkes process framework. We then apply this LOB modeling approach to a state-of-the-art deep reinforcement learning (RL) algorithmic and HFT problem, a relatively new but more robust method for analyzing such mathematical optimization problems. We model 12 types of LOB events, capturing a significant portion of real-world LOB activity. This can be extended to include additional event types available in certain markets, provided their impact on the midprice process and the intensity function governing event frequencies is well understood. Some of the more recent literature has made similar attempts, such as in \cite{gavsperov2022deep} where they devised a MM framework that applies a midprice model that evolves based on the linear MVHP model formulated in \cite{lu2018high}. However, we believe that creating a MVHP that performs a nonlinear transformation to a Linear MVHP to be more general, and more in line with how LOB pricing data should be modeled. Empirical evidence supporting this theory is provided in \cite{lu2018high} and \cite{shi2022state}, demonstrating that a nonlinear MVHP model fits LOB data better than a linear MVHP or other traditional Hawkes models.

Building on this LOB event structure, we construct a midprice process using a neurally self-modulated multivariate point process, known as a Neural Hawkes Process (Neural HP), which was first introduced in \cite{mei2017neural}. To date, we have identified only two applications of the Neural HP in modeling LOBs, as also noted in the recent survey by \cite{jain2024limit}. In \cite{shi2022state}, empirical evidence suggests that their LOB event model, based on four LOB event types, better simulates LOB data compared to traditional stochastic HP models and other deep learning methods. In \cite{kumar2024deep}, the Neural HP is used to simulate LOB data after fitting it to specific datasets while also incorporating trader actions through an agent-based model. In addition to capturing the advantages of a nonlinear MVHP, which enables more accurate modeling of self- and cross-excitation relationships between LOB events, Neural HPs also address some of the limitations still present in traditional HP models. These limitations include the requirement for the excitation function to generate only positive values, the additive nature of past events, and the fixed decay structure of the intensity function. For example, within LOB data it is clear from the empirical evidence in \cite{lu2018high} that inhibition effects, rather than excitation effects, are quite common between certain LOB events, which a well-trained Neural HP model can account for. Consequently, Neural HP models are better suited for capturing these complex dependency relationships. 

The main contributions of this paper can be summarized as follows:
\begin{enumerate}
    \item We develop a Neural Hawkes LOB simulation framework based on 12 key LOB events observed in electronic financial markets. Empirical evidence demonstrates how this event-based approach closely aligns with real-world LOB event arrivals, capturing their nonlinear dependencies. This is a crucial feature for HFT, as individual events or sequences of events carry granular information essential to short-term traders.
    \item This event-based LOB structure allows us to develop an asset price simulation driven by a unique jump process, where jumps correspond to the probability of specific jump sizes occurring within each LOB event category. We demonstrate how this approach can be structured based on asset-specific event characteristics and applied to various HFT scenarios. 
    \item This is the first work to apply a Neural Hawkes event-based LOB model to a High-Frequency MM strategy and demonstrate how this structure enables more realistic HFT strategy backtesting. We believe this represents a significant extension of previous MM literature, as it allows for more accurate simulation of trade order fills based on the LOB events required to complete a trade. Additionally, it improves the modeling of LOB event frequencies and timing. These aspects are crucial for High-Frequency MM simulations and have received limited attention in existing research. 
\end{enumerate}

The rest of this paper will proceed as follows. In Section 2, we introduce the midprice model used throughout the paper. Section 2.1 provides a discussion and analysis of the included LOB events, while Section 2.2 presents an overview of the nonlinear MVHP modeling framework. In Section 3, we describe the Neural MVHP midprice simulation framework. Section 3.1 introduces the Neural Hawkes process model, while Section 3.2 details the midprice simulation process and discusses key simulation results. In Section 4, we will then proceed to provide an application under a Deep RL framework, where we will focus on a High-Frequency MM trading problem. Here, we conduct a comparative analysis between the simulated and real LOB data across five different assets. Lastly, we will discuss our conclusions and future research recommendations.   

\section{Midprice Process Modeling}
In this section we proceed to discuss the formulation of our Midprice modeling process, under a nonlinear MVHP framework. Over the last 15-20 years, HPs have seen many applications in finance, where a broad survey can be found in \cite{bacry2015hawkes}, and more recently also in Algorithmic and HFT. A HP is a self-exciting point process where the occurrence of an event increases the likelihood of future events occurring within a short time frame. This makes it well-suited for modeling clustered event arrivals, such as LOB activity in financial markets. Examples of applications in the LOB modeling space include \cite{abergel2015long} where they study the long-term behavior of applying HP to LOBs, \cite{cartea2018algorithmic} where a order flow metric is modeled using a HP, \cite{swishchuk2020general} and \cite{swishchuk2019compound} which formulates a selection of General Compound HP models that can be applied within certain trading problems, where different applications can be found in \cite{roldan2022optimal} and \cite{lalor2024algorithmic}. This is by no means an extensive review of HP applications in LOBs, for this we refer the reader to the survey paper by \cite{jain2024limit}. However, it is clear from much of the recent literature to date that HPs are a suitable choice for modeling LOB events, particularly from a point process standpoint. 

The rest of this section will proceed as follows. In Section 2.1, we provide an overview of the 12 LOB events examined in this paper, along with an analysis of their occurrence frequencies in real LOB data. In Section 2.2, we will describe the nonlinear MVHP dynamics used to model these events, along with our midprice process that represents price moves based on these events. 

\subsection{\secondlevelheading{Limit Order Book Events}}

In selecting which LOB events to model, we follow a similar set of events as used in \cite{lu2018high}, \cite{law2019market} and \cite{gavsperov2022deep}. In these works, the focus is on how each respective LOB event affects the midprice i.e., whether these events caused the midprice to increase, decrease or remain unchanged. Irrespective of their effect on the midprice, they all affect the MVHP intensity function that will be introduced in Section 2.2, through the self- and cross-excitation effects that can be modeled by nonlinear MVHP models. Subsequently, these LOB events can be introduced as follows: 

\begin{enumerate}
\item Aggressive Limit Buy Order ($LB^+$): A limit order that moves the midprice up by posting at a higher bid price than the previous best bid. 
\item Aggressive Limit Sell Order ($LS^-$): A limit order that moves the midprice down by posting at a lower ask price than the previous best ask. 
\item Aggressive Market Buy Order ($MB^+$): A market order that moves the midprice up and depletes at least one ask queue. 
\item Aggressive Market Sell Order ($MS^-$): A market order that moves the midprice down and depletes at least one bid queue.
\item Aggressive Limit Buy Cancellation ($BC^-$): A canceled buy limit order that leads to a decrease in the midprice as the queue size at the previous best bid price is emptied.
\item Aggressive Limit Sell Cancellation ($SC^+$): A canceled sell limit order that leads to an increase in the midprice as the queue size at the previous best ask price is emptied.
\item Non-aggressive Limit Buy Order ($LB^0$): A limit buy order that leaves the midprice unchanged. 
\item Non-aggressive Limit Sell Order ($LS^0$): A limit sell order that leaves the midprice unchanged.
\item Non-aggressive Market Buy Order ($MB^0$): A market buy order that leaves the midprice unchanged. 
\item Non-aggressive Market Sell Order ($MS^0$): A market sell order that leaves the midprice unchanged.  
\item Non-aggressive Limit Buy Cancellation ($BC^0$): A limit buy order cancellation that leaves the midprice unchanged. 
\item Non-aggressive Limit Sell Cancellation ($SC^0$): A limit sell order cancellation that leaves the midprice unchanged. 
\end{enumerate}

In this paper, we used \cite{LOBSTER} data, which is a website partnered with Nasdaq and the Universität Wien that offers high-frequency data on multiple stocks, upon which many researchers now rely on in algorithmic and HFT research. They offer some free data on one day, June 21, 2012, throughout the full trading day (9:30 AM–4:30 PM EST). This free data includes LOB data up to 10 price levels, along with the transaction message feeds, for stocks listed on the NASDAQ exchange. Table \ref{tab:LOB_Event_Counts} presents the frequency of each previously described LOB event across five stock tickers (Apple-AAPL, Amazon-AMZN, Alphabet-GOOG, Intel Corp.-INTC, and Microsoft-MSFT), which were derived using all 10 LOB levels. The last column of Table \ref{tab:LOB_Event_Counts} displays the probability of each LOB event occurring, conditional on the occurrence of any LOB event, based on data from these five stocks. 

\begin{table}[h]
    \centering
    \begin{tabular}{lcccccc}
        \toprule
        & \multicolumn{5}{c}{Stock Ticker} & \\ 
        \cmidrule(lr){2-6}
        Event-Type & AAPL & AMZN & GOOG & INTC & MSFT & Probability \\
        \midrule
        $LB^+$  & $16,805$  & $6,611$  & $6,496$  & $739$  & $951$  & $0.01497$ \\
        $LS^-$  & $17,474$  & $6,993$  & $6,381$  & $869$  & $1,078$  & $0.01554$ \\
        $MB^+$  & $5,876$  & $1,871$  & $1,748$  & $725$  & $931$  & $0.00528$ \\
        $MS^-$  & $6,227$  & $2,368$  & $2,111$  & $695$  & $888$  & $0.00582$ \\
        $BC^-$  & $8,969$  & $5,013$  & $3,904$  & $95$  & $104$  & $0.00857$ \\
        $SC^+$  & $8,999$  & $4,701$  & $3,444$  & $94$  & $108$  & $0.00822$ \\
        $LB^0$  & $65,714$ & $56,792$ & $27,626$ & $162,421$ & $153,972$ & $0.22101$ \\
        $LS^0$  & $91,021$ & $61,588$ & $30,755$ & $140,761$ & $173,565$ & $0.23577$ \\
        $MB^0$  & $12,385$ & $3,644$ & $4,121$ & $17,939$ & $15,784$ & $0.02552$ \\
        $MS^0$  & $10,502$ & $3,535$ & $3,697$ & $13,123$ & $15,811$ & $0.02211$ \\
        $BC^0$  & $65,830$ & $56,413$ & $27,839$ & $153,903$ & $144,480$ & $0.21245$ \\
        $SC^0$  & $90,588$ & $60,248$ & $29,793$ & $132,675$ & $161,092$ & $0.22474$ \\
        \bottomrule
    \end{tabular}
    \caption{The total number of occurrences of each LOB event type from up to ten LOB levels in the LOBSTER dataset on June 21st, 2012, along with the probability of each event occurring based on these 5 stocks.}
    \label{tab:LOB_Event_Counts}
\end{table}

As it is clear that LOBs evolve based on the above type of LOB events, we believe it is paramount that any midprice simulation process account for this, specifically when studying high-frequency data and short-term trading strategies. As in \cite{lu2018high} and \cite{gavsperov2022deep}, we will proceed to show how each of these LOB events can be modeled using a nonlinear MVHP model, as MVHPs have been commonly used to model the relationship between these events. In the HFT strategy simulation literature, only a Linear MVHP has been used to simulate the LOB data, thus we aim to extend previous work by simulating a nonlinear MVHP. Nonlinear MVHPs offer greater flexibility in modeling complex dependencies between events, as they allow for non-additive, state-dependent excitation effects and can capture inhibition or saturation effects that linear models cannot. This makes them better suited for applications like LOB modeling, where event interactions are highly dynamic and not strictly additive.  

\subsection{\secondlevelheading{Multivariate Hawkes Process Midprice Modeling}}

Consider $(N(t)=(N_1(t), N_2(t),..., N_m(t))$ to be a simple multivariate counting process, where only one event can occur at any given time. This assumption is realistic for LOB data due to its high-resolution nature, where each LOB event has a unique timestamp, almost surely. The conditional intensity process of this simple multivariate counting process is $\lambda(t)=(\lambda_1(t), \lambda_2(t),..., \lambda_m(t))$, representing the number of LOB events of type $i$ occurring up to time $t$, for $i = 1,...,m$. In our case $m=12$, corresponding to the number of different LOB event types. The intensity function for each $\lambda_i(t)$, for the $i$-th event at time $t$, can be defined as follows:
\begin{align}
\lambda_i(t) = \phi_i\left(\lambda_i+\int_{(0,t)}\sum_{j=1}^m\mu_{ij}(t-s)dN_j(s)\right),
\label{eq:hawkes_intensity}
\end{align}
where $\lambda_i$ is the baseline intensity of the $i$-th event which may be time dependent, $\mu_{ij}$ is a kernel function (often an exponential or power-law function in the literature due to their simplicity) describing the influence of past events of type $j$ on type $i$ events, and $\phi_i$ is a nonlinear function that transforms the linear MVHP into a nonlinear MVHP. For our purposes, we will select the exponential kernel function for simplicity, and so we set $\mu_{ij} = \alpha_{ij}e^{-\beta_{ij}t}$. Equation (\ref{eq:hawkes_intensity}) can then be rewritten as follows:
\begin{align}
\lambda_i(t) = \phi_i\left(\lambda_i+\int_{(0,t)}\sum_{j=1}^m\alpha_{ij}e^{-\beta_{ij}(t-s)}dN_j(s)\right),
\label{eq: hawkes_intensity_exp} 
\end{align}
where $\alpha$ and $\beta$ govern the self($i=j$)- and cross($i \neq j$)- excitation effects between each of the events being modeled.   

Next, we divide our list of LOB events in Section 2.1 into three categories, based on their respective midprice movements. Notice the shorthand notation we gave these events in Section 2.1, where the exponents $+$, $-$, and $0$ refer to LOB events that lead to an increase, a decrease, or no change in the midprice. These LOB event categories can be described as follows.
\begin{itemize}
\item ${O}_{u}=\{MB^+,LB^+,SC^+\}$ represents the LOB events that cause the midprice to increase. 
\item ${O}_{d}=\{MC^-,LS^-,BC^-\}$ represents the LOB events that cause the midprice to decrease. 
\item ${O}_{n}=\{MB^0,MS^0,LB^0,LS^0,BC^0,SC^0\}$ represents the LOB events that do not cause the midprice to change. 
\end{itemize} 
Note also that the LOB events in the sets $O_u$ and $O_d$ are referred to as aggressive events since they cause the midprice to change, whereas events in the set $O_n$ are referred to as non-aggressive events as they cause no change in the midprice. 

As the LOB evolves based on these types of events occurring discretely, we now define the evolution of a midprice process in discrete-time as follows:
\begin{align}
\begin{split}
    V(t+\Delta t) &= V(t) + \frac{\Delta}{2}\times\left[\sum_{k\in O_u} \mathbb{I}_k(t)a(X_{k}) - \sum_{l\in O_d} \mathbb{I}_l(t)b(X_{l}) \right. \\
    &\quad \left. + \sum_{m\in O_n} \mathbb{I}_m(t)c(X_{m})\right]  \\
    &= V(t) + \frac{\Delta}{2}\times\left[\sum_{k\in O_u} \mathbb{I}_k(t)a(X_{k}) - \sum_{l\in O_d} \mathbb{I}_l(t)b(X_{l})\right].
\end{split}
\label{eq: midprice_process}
\end{align}
Note that the third term vanishes because the jump sizes are always zero for the non-aggressive LOB events present in the set $O_n$. The right-hand side of Equation (\ref{eq: midprice_process}) can be summarized as follows:
\begin{itemize}  
    \item $V(t)$ represents the value of the asset at time $t$.  
    \item $\Delta t$ represents the time step, which here is non-uniform as LOB events occur at irregular time intervals. 
    \item $\Delta$ represents the tick size/bid-ask spread of the asset being modeled, where $\frac{\Delta}{2}$ represents the half-spread. 
    \item $\sum_{k\in O_u} \mathbb{I}_k(t)a(X_{k})$ accounts for the upward movements (represented by the LOB events in the set $O_u$), where,
    \begin{itemize}
        \item $\mathbb{I}_k(t)$ checks if an event $k$ occurs or not at time $t$,
        \item $a(X_k)$ describes the mapping from the current state of the asset to a price move, where $X_k$ 
        represents the type of price movement. This price movement can be modeled where there are many possible tick movements i.e., 1 tick, 2 ticks, 3 ticks, etc. 
    \end{itemize}   
    \item $\sum_{l\in O_d} \mathbb{I}_l(t)b(X_{l})$ and $\sum_{m\in O_n} \mathbb{I}_m(t)c(X_{m})$ can be described similarly for downward and non-change movements, respectively. 
    \item Recall that events in the category ${O}_{u}(t)$ (${O}_{d}(t)$) cause the midprice to increase (decrease), while events in the category ${O}_{n}(t)$ cause no change in the midprice, but still affect the intensity function in Equations (\ref{eq:hawkes_intensity})-(\ref{eq: hawkes_intensity_exp}). 
\end{itemize}

This is a more general version of the mid-price process given in \cite{lu2018high}, where they assumed that the jump size for each $+$ and $-$ event is always one tick. If we simplified our model to their model, the jump size for each $+$ or $-$ LOB event would be equal to $1$, that is, $a(X_{k})=b(X_{k})=1$ at each time step. This would lead us to get the following.

\begin{align}
    V(t+\Delta t) &= V(t) + \frac{\Delta}{2}\times\left[\sum_{k\in O_u} \mathbb{I}_k(t) - \sum_{l\in O_d} \mathbb{I}_l(t)\right]
    \label{eq: lu_midprice_process}
\end{align}
which is the same as midprice process as in \cite{lu2018high}, but in discrete-time.

\section{Midprice Simulation via a Neural Hawkes Process}

The Neural HP extends traditional HP's by formulating \sloppy continuous-time recurrent neural networks (RNNs) to model event sequences with adaptive temporal dependencies, as are apparent in nonlinear MVHP models. This approach overcomes many of the limitations that appear in most of the classical HP models, such as assuming a fixed parametric form for the intensity function. The Neural HP model in the original paper, by \cite{mei2017neural}, is built using a Long Short-Term Memory (LSTM) RNN, which learns the intensities of specific events over time. The idea of modeling LOBs based on the structure of a Neural HP is relatively new, and it aims to extend previous work on applying HPs to LOB models by allowing the process to capture more real-world phenomena. In particular, inhibition and inertia effects in LOB data, as shown in \cite{lu2018high}, can now be modeled. More specifically, the $\alpha_{i,j}$ and $\lambda_i$ values in Equation \eqref{eq: hawkes_intensity_exp} can now be negative, which is not possible in the more restrictive nonlinear MVHP framework. This could result in a nonnegative activation function, i.e., negative intensities, which subsequently gets passed through a nonlinear transfer function to ensure positivity. Additionally, the combined effect of past events is not required to be additive, and it can also account for delayed effects that are often not apparent through the simple decay constant used under regular HPs like in Equation (\ref{eq: hawkes_intensity_exp}). Thus, the Neural Hawkes process has the ability to overcome some of the shortcomings in traditional HP model, where now the influence of past events on future events can be greater than, less than, or even reduce their impact, and it may also depend on the order in which past events occurred. 

In this section, we will proceed by introducing our Neural HP model in Section 3.1, where we will outline the network architecture and how we conducted our training and testing, where a summary of these results will also be given. We will also show how the 12 LOB events in our model can be simulated based on these results. Then in Section 3.2, we will discuss how we use these results to simulate our LOB event-based midprice process, which extends much of the previous literature where an independence between events and price processes is often assumed. We will also provide sample midprice simulations for the assets in the LOBSTER data, where we provide an analysis of how many parts of these results are in line with the real LOB data discussed in Section 2.1.    

\subsection{\secondlevelheading{Neural Hawkes Process Framework}}
\begin{figure}[H]
	\centering
	\includegraphics[width=0.5\linewidth]{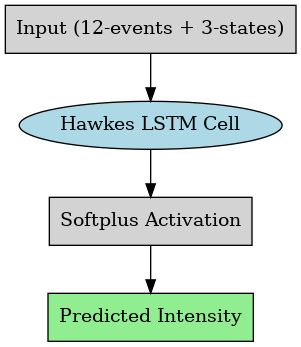}
	\caption{A visualization of the structure of our LSTM Neural Hawkes LOB model. }
	\label{fig:NeuralHawkes}
\end{figure}

Here we explain the main components of the Neural HP model we developed, where many parts are heavily in line with the standard Neural Hawkes model in \cite{mei2017neural} and in \cite{shi2022state}, where the latter is an extension of the original Neural Hawkes model applied to a state-dependent LOB model. One can also visualize the structure of the network simulation process we created in Figure \ref{fig:NeuralHawkes}. To give a brief overview before going into more depth, this network receives the LOB event-type and a market state variable as an input, the LSTM cell then approximates the structure of the intensity function which evolves via the exponential kernel, and as an output, the network receives a value for each LOB event's new intensity. Our model is trained using the LOBSTER data described earlier, where we split the data into 60\% training, 20\% validation, and 20\% testing. In the following parts of this section, we will describe the working parts of this framework in more detail.  

The main working parts of this framework can be summarized as follows:
\begin{itemize} 
    \item Input: The model takes as input the LOB event type through one-hot encoding of event-type and a market state, which is an additional volume imbalance indicator feature representing the current state of the market, as first described in \cite{cartea2018enhancing} and also used in \cite{shi2022state}. This market state feature uses the volume posted on the best bid or best ask as a proxy for predicting short-term order flow. More specifically, denote this predictor as $I$ where: 
    \begin{align}
        I = \frac{v^{b(1)}-v^{a(1)}}{v^{b(1)}+v^{a(1)}},
        \label{eq: short_term_OF}
    \end{align}
    where $v^{b(1)}$ and $v^{a(2)}$ is the volume posted on the best bid and ask, respectively. Then, the state of the market is defined categorically as follows:
    \begin{align}
        x =
            \begin{cases} 
            0, & \text{if } I \in [-1, \theta] \\ 
            1, & \text{if } I \in [-\theta, \theta] \\ 
            2, & \text{if } I \in [-\theta, 1] 
            \end{cases},
            \label{eq: mkt_state}
    \end{align}
    where here $x$ denotes the market state and $\theta$ is a fixed parameter setting the boundary values. In our model, we set $\theta = 0.4$ as in \cite{shi2022state}. $\theta$ essentially determines the boundary for specifying the difference between a balanced (1) and unbalanced market (0 or 2). This predictor essentially states that if the volume imbalance indicator predicts an up movement, events which cause the midprice to go up are more likely to occur, with a similar logic for the down and no move categories. 
    \item LSTM Memory Update: The famous LSTM structure, as first introduced in the seminal paper by \cite{hochreiter1997long}, is used to maintain and update a hidden state $h_t$ and a cell state $c_t$. Here, we follow the method in \cite{shi2022state} under our more granular LOB event space, where they stack $m$ Continuous-Time LSTM units for encoding the input information, where recall $m$ is the number of LOB event types. This extension to the traditional LSTM structure in \cite{mei2017neural} allows the hidden state parameters to now be specifically computed for each of our LOB event types, where the standard approach previously shared these parameters. The general update mechanism for this LSTM system, as used in our analysis, can be described as follows:
    \begin{align}
        i_t &= \sigma(W_i [x_t, h_{t-1}] + b_i) && \text{(Input Gate)} \label{eq:input_gate} \tag{7a} \\
        f_t &= \sigma(W_f [x_t, h_{t-1}] + b_f) && \text{(Forget Gate)} \label{eq:forget_gate} \tag{7b} \\
        g_t &= f_t \odot c_{t-1} + i_t \odot \tanh(W_g [x_t, h_{t-1}] + b_g) && \text{(Target Cell State)} \label{eq:candidate_cell} \tag{7c} \\
        \delta_t &= \exp\{W_{\delta}[x_t,h_{t-1}] + b_\delta\} && \text{(Decay Rate)} \label{eq:decay_rate} \tag{7d} \\
        c_t &= g_t + (c_{t-1}-g_t)\odot \exp \{-\delta_t\Delta t\} && \text{(Cell State)} \label{eq:cell_state} \tag{7e} \\
        o_t &= \sigma(W_o [x_t, h_{t-1}] + b_o) && \text{(Output Gate)} \label{eq:output_gate} \tag{7f} \\
        h_t &= o_t \odot \tanh(c_t) && \text{(Hidden State)} \label{eq:hidden_state} \tag{7g}
    \end{align}

    where:  
    \begin{itemize}
        \item $i_t$, $f_t$, and $o_t$ are the input, forget, and output gates. Intuitively, the input gate determines how much of the new target cell state should be added to the memory cell, the forget gate decides how much of the previous cell state should be retained or forgotten, and the output gate is then a filtered version of the memory cell state.
        \item $g_t$ is the target cell state, which is the update proposed to the memory cell based on the current input and the previous hidden state. 
        \item $c_t$ is the memory cell, which stores long-term information by combining the previous memory cell and the new target cell state.
        \item $h_t$ is the hidden state, which is the output of the LSTM cell, which is passed to the next time step as in a regular RNN.
        \item $x_t$ is the input at time $t$, which includes information about the type of event, the market state, and the time step.
        \item The activation functions, $\sigma$ and $\tanh$, represent the sigmoid and hyperbolic tangent functions, respectively.
        \item $\odot$ represents the element-wise Hadamard product of two matrices. 
    \end{itemize} 
    \item Output (LOB Event Intensity Function): Based on the LSTM RNN structure described, the intensity function in Equation (\ref{eq: hawkes_intensity_exp}) can now be reformulated as follows: 
    \begin{align}
        \lambda_i(t) = \phi_i(h_i(t)) = ln\left(1+e^{h_i(t)}\right).
    \end{align}
    Similarly to Equation \eqref{eq: hawkes_intensity_exp}, each $\lambda_i(t)$ jumps discontinuously and drifts towards a stable baseline value $\lambda_i$, but in the Neural Hawkes process setup, however, these dynamics are governed by a hidden state vector $h(t)\in (-1, 1)^D$ as defined in Equation \eqref{eq:hidden_state}, where each events intensity $\lambda_i(t)$ is determined by the corresponding component $h_i(t)$. The influence of each LOB event on the intensity function is captured by the memory cell vector $c(t)$ in the LSTM RNN. More specifically, the LSTM input and forget gates defined in Equations \eqref{eq:input_gate}-\eqref{eq:forget_gate} determine how past event are remembered or forgotten. The input gate determines how much influence new events have on the cell state, similar to how $\alpha_{ij}$
    scales the contribution of past events in a Hawkes process but in a more dynamic way. The decay function in Equation \eqref{eq:decay_rate} is learned dynamically, allowing it to adapt to different contexts, unlike the fixed decay $\beta_{ij}$ in a standard HP. Meanwhile, the forget gate determines how much of the past memory is preserved before decay is applied. Lastly, it is clear that $\phi$ is represented by the Softplus function, which is used to perform the nonlinear transformation, and this ensures that each LOB event's predicted intensity function is nonnegative. Due to the $m$-stacked LSTM structure, each LOB event now also has a unique intensity based on its specific latent dynamics.  
    \item Training Phase: Here, the model is trained over $20$ epochs, where we use batch sizes of $256$ and a rolling window approach over each time step and sequence lengths of $100$. As in \cite{shi2022state}, our loss function will primarily focus on the loss arising from the Hawkes dynamics, which is defined by the following negative log-likelihood function,
    \begin{align}
        \mathcal{L} = \sum_{j=1}^{J-1}\sum_{i=0}^{m} log(\lambda_i(t_{j+1}))-\int_{t_j}^{t_{j+1}}\lambda_i(s)ds,
        \label{eq: loss_func}
    \end{align}
    over the likelihood of event times $t_1, t_2, ..., t_J$, where $J$ represents the total number of event times. The parameters are then updated using the standard RMSprop (Root Mean Square Propagation) adaptive gradient-based optimization algorithm, where a learning rate of $0.002$ is used. The accuracy rate over each batch is also tracked for evaluation. 
\end{itemize}

\begin{table}[H]
    \centering
    \renewcommand{\arraystretch}{1.2} 
    \begin{tabular}{lccccc}
        \toprule
        & AAPL & AMZN & GOOG & INTC & MSFT \\
        \midrule
        \textbf{Training} & & & & & \\
        \hspace{1em} Loss & 1.7739 & 1.6952 &2.6181 & -0.4972 & -0.6151 \\
        \hspace{1em} Accuracy & 0.4588 &0.4401 &0.3778 & 0.5675 & 0.5489  \\
        \midrule
        \textbf{Testing} & & & & & \\
        \hspace{1em} Loss &1.5426 &1.5662 &2.76 & -0.2038 & -0.5639 \\
        \hspace{1em} Accuracy & 0.4054 & 0.3904 &0.3153 & 0.4969 & 0.5124 \\
        \bottomrule
    \end{tabular}
    \caption{Training and Testing Results in AAPL, AMZN, GOOG, INTC and MSFT from the LOBSTER sample datasets. }
    \label{tab:train_test_results}
\end{table}

\begin{figure}[H]
	\centering
	\includegraphics[width=0.95\linewidth]{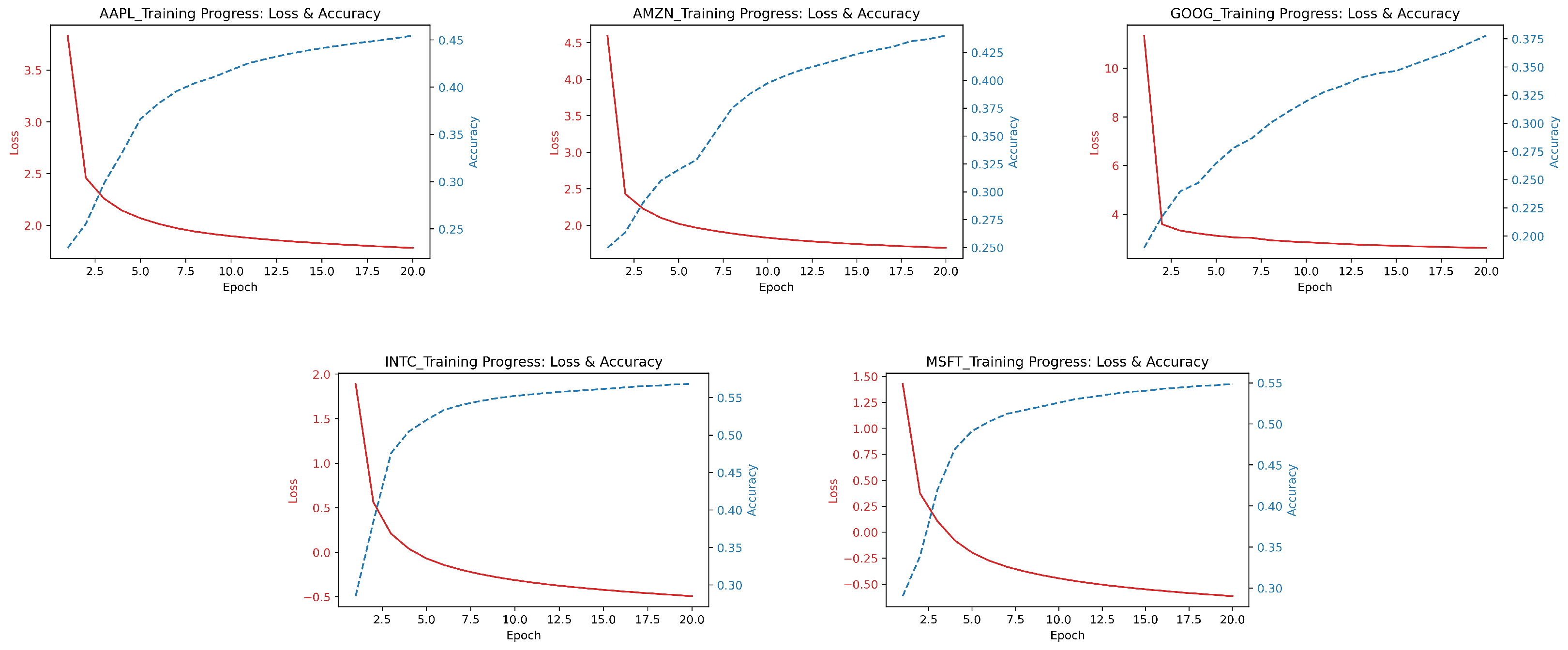}
	\caption{A visualization of the loss function and accuracy during the training phase in AAPL, AMZN, GOOG, INTC and MSFT, where the loss (red) and the accuracy (blue) is shown over 20 epochs.}
	\label{fig:train_loss_acc}
\end{figure}

In Table \ref{tab:train_test_results}, we show the results for the training and test sets in the 5 assets studied (AAPL, AMZN, INTC, GOOG, MSFT). Here, we show the value of the loss function, as computed using Equation (\ref{eq: loss_func}), and the accuracy rate, which measures the probability of correctly predicting the next event in the batch. Table \ref{tab:train_test_results} shows us that the training results generalize to the testing results quite well in this setting. In Figure \ref{fig:train_loss_acc}, we also provide a visualization of the training results in all five assets, where we show how the loss function and accuracy evolved over the first 20 epochs. It is clear that training improved these measures as the loss decreased and the accuracy increased, and that the generalization error from training to testing is relatively low in this setting. As can be seen by the shape of loss function, it is clear that after 20 epochs more training would likely not improve the results significantly and would more likely just lead to an overfit model. It is clear that the best results were in INTC and MSFT, which coincidentally is also the case in the results in \cite{shi2022state} for their model with 4 LOB events. 

Now, based on the results of the training phase, the 12 LOB events in our study can be simulated. Here, we adopt the widely used approach for simulating MVHPs: Ogata's thinning algorithm, as first derived in \cite{lewis1979simulation}, is a rejection sampling method used to simulate non-homogeneous Poisson processes. Ogata's thinning algorithm was first applied to simulating MVHPs in \cite{liniger2009multivariate} and was also used for simulation purposes in the original Neural Hawkes paper by \cite{mei2017neural}, as well as in \cite{shi2022state} for simulating their set of four LOB events. The algorithm first simulates an inhomogeneous Poisson process with an upper-bound intensity and then selectively accepts events based on its conditional intensity function. The waiting time, $\Delta t$, follows an exponential distribution, that is, $\Delta t \sim Exp(\Lambda_t)$, where here $\Lambda_t = \sum_{i=1}^m\lambda_i(t)$, and $m=12$ represents the number of LOB events. This has the probability density function $P(\Delta t) = \Lambda(t)e^{-\Lambda (t) \Delta t }$, where the expected time until the next event is $\mathbb{E}[\Delta t] = \frac{1}{\Lambda (t)}$. This is then sampled using the standard method $\Delta t = -\frac{log(U)}{\Lambda(t)}$, where $U \sim Uniform (0,1)$. The probability of each type of LOB event $i$ occurring is then modeled as,
\begin{align}
    P(i|LOB \ event \ at \ t) = \frac{\lambda_i(t)}{\Lambda(t)}.
\end{align}
Intuitively, it is now quite obvious that LOB events with a higher intensity are more likely to be chosen. 

\begin{table}[h]
    \centering
    \begin{tabular}{lccccc}
        \toprule
        & \multicolumn{5}{c}{Stock Ticker} \\
        \cmidrule(lr){2-6}
        Event-Type & AAPL & AMZN & GOOG & INTC & MSFT \\
        \midrule
        $LB^+$  & $2,790$ & $1,984$ & $2,781$ & $1,884 $ & $2,290$ \\
        $LS^-$  & $3,092$ & $1,640$ & $2,841$ & $1,836 $ & $2,239$ \\
        $MB^+$  & $3,111$ & $1,764$ & $2,625$ & $1,937 $ & $2,534$ \\
        $MS^-$  & $2,469$ & $1,572$ & $2,639$ & $1,966$ & $2,211$ \\
        $BC^-$  & $2,265$ & $1,532$ & $2,675$ & $1,852$ & $2,241$ \\
        $SC^+$  & $2,451$ & $1,773$ & $2,430$ & $1,917$ & $2,314$ \\
        $LB^0$  & $4,417$ & $7,953$ & $8,324$ & $7,998$ & $5,917$ \\
        $LS^0$  & $7,653$ & $11,008$ & $7,235$ & $8,788$ & $8,336$ \\
        $MB^0$  & $4,173$ & $1,954$ & $2,866$ & $1,878$ & $5,783$ \\
        $MS^0$  & $2,806$ & $1,488$ & $3,543$ & $4,694$ & $2,359$ \\
        $BC^0$  & $3,663$ & $5,711$ & $5,925$ & $6,474$ & $8,380$ \\
        $SC^0$  & $12,310$ & $12,821$ & $7,316$ & $9,976$ & $6,587$ \\
        \bottomrule
    \end{tabular}
    \caption{The total number of occurrences of each LOB event type over 200 simulations in AAPL, AMZN, GOOG, INTC, MSFT.}
    \label{tab:LOB_Event_Counts_Sim}
\end{table}

\begin{figure}[H]
	\centering
	\includegraphics[width=0.95\linewidth]{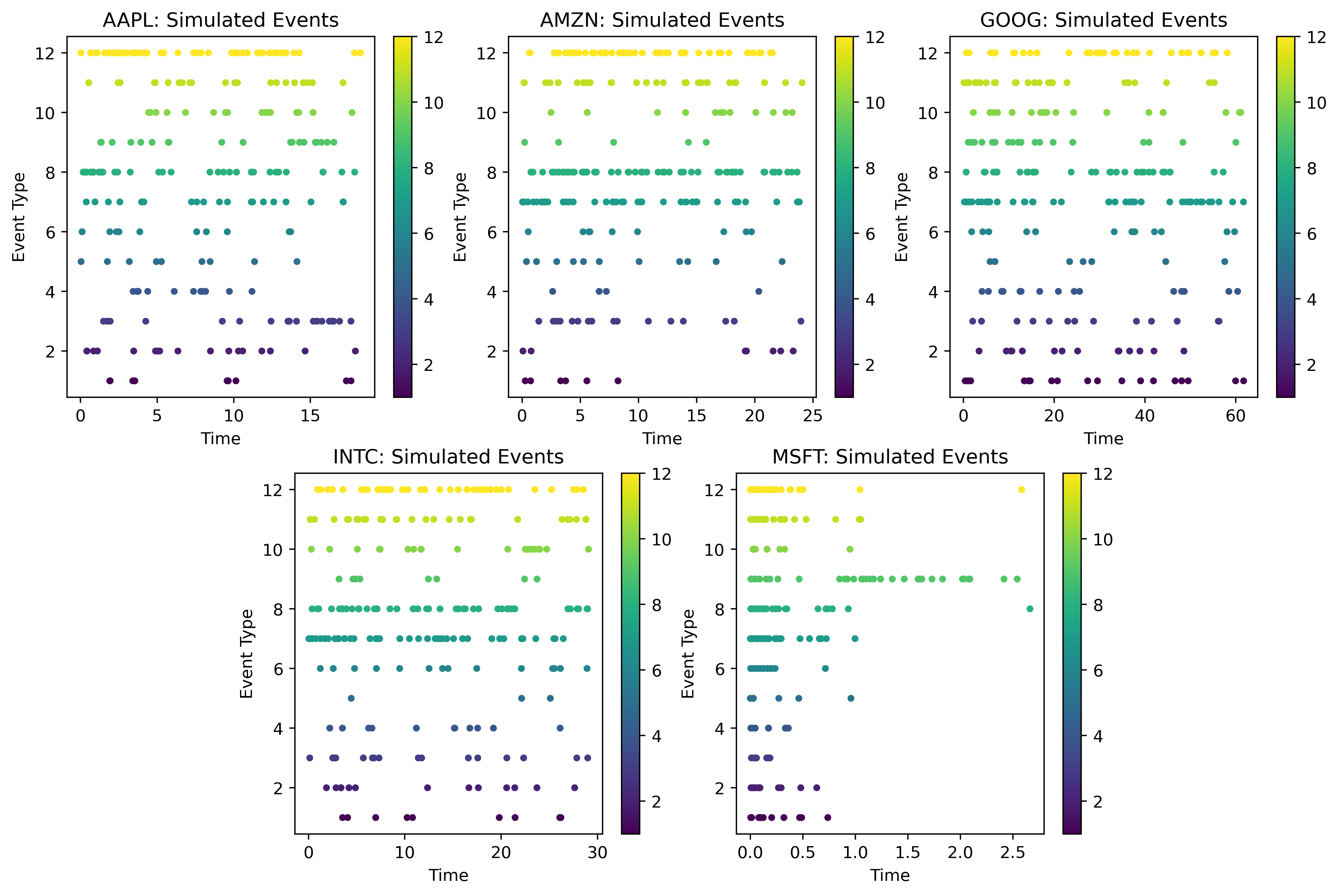}
	\caption{A visualization of how often each LOB event occurred over the first of the two hundred simulations conducted in AAPL, AMZN, GOOG, INTC and MSFT, where each color-coded dot represents the occurrence of a specific event. The y-axis values, ordered from 1 to 12, are in the same order as the events in Table \ref{tab:LOB_Event_Counts_Sim}.}
	\label{fig:LOB_Event_Int_Sim}
\end{figure}

In our setting, we ran 200 simulations over each batch. Table \ref{tab:LOB_Event_Counts_Sim}, above, presents the cumulative frequency of each LOB event in the simulation, while Figure \ref{fig:LOB_Event_Int_Sim}, below, provides a visualization of their occurrence over time in the first simulation. The results in Table \ref{tab:LOB_Event_Counts_Sim} show that non-aggressive events (with exponent $0$), occur more often than aggressive events (with exponent $+/-$), which is in line with the real data in Table \ref{tab:LOB_Event_Counts}. While the simulation still exhibits a higher proportion of aggressive events relative to non-aggressive events compared to the real data, the overall event distribution remains informative for analysis. Thus, it is clear that the model has learned which events are more likely to occur, but still with some inaccuracies, as also clearly shown from the training and testing results in Table \ref{tab:train_test_results}. Note, however, that this type of Neural Hawkes model has been proven to beat the standard benchmarks of traditional Hawkes models in \cite{shi2022state}, thus we certainly still believe this model to be an improvement on many of the previous Hawkes LOB models in the literature, particularly given its event-based structure. 

\subsection{\secondlevelheading{Midprice Simulation Process}}

In mathematical finance, midprice processes (often referred to simply as price processes) are typically simulated by discretizing continuous-time differential equations that incorporate diffusion, pure jump, jump-diffusion, or other pricing model dynamics. Some famous examples of models used to simulate asset prices include using an Euler discretization of the famous Geometric Brownian Motion model based on \cite{black1973pricing}, which may also include jumps as in \cite{merton1976option}, then Euler-Maruyama or Milstein methods are also often used (see \cite{kloeden1992stochastic}, as well as many other practical formats via Monte Carlo simulation. In much of the Algorithmic and HFT literature, similar simulation techniques are commonly employed. The most widely used approaches, as discussed in the recent survey paper by \cite{jain2024limit}, include point process-based models where zero-intelligence models like Poisson processes and basic Hawkes process models with dependency structures are the most popular. This survey paper also touches on agent-based models and deep learning-based simulation models. However, most of these simulation techniques overlook the event-based nature of LOB data, which limits their effectiveness in simulating HFT strategies, where granular LOB information is crucial for making informed decisions. We found a previous study, \cite{gavsperov2022deep}, that simulates an event-based LOB model within a high-frequency market-making (MM) strategy framework using a Linear MVHP. As noted earlier, this approach has certain limitations that the Neural Hawkes model aims to overcome. In the high-frequency MM domain, an agent-based model has also been developed, as shown in \cite{kumar2024deep}, where a Neural Hawkes model is used to capture interactions between different trading agents in the market. Our work, however, focuses on the distinct event-based structure of LOB data, which can be difficult for an agent-based model to fully capture without modeling a broad range of agent behaviors at a highly granular level.

To simulate the midprice process, we use Equation (\ref{eq: midprice_process}) as defined in Section 2.2. Here, it is clear that the process must start with an initial midprice V(0), where we will use the first midprice value in the data. Each subsequent midprice movement is then determined by our event-based intensity function, computed through our Neural Hawkes simulation process. More simply, we can redefine Equation (\ref{eq: midprice_process}) in a simpler form as needed for a simulation. Here, at each iteration, the price at $V(t+\Delta t)$ is calculated as follows,
\begin{align}
    V(t+\Delta t) = V(t) + \operatorname{sgn}(\Delta V(t)) |\Delta V(t)|
\end{align}
where recall $\Delta t$ is a non-uniform time step based on the interrarrival times of the LOB events. Here, $\operatorname{sgn}(\Delta V(t))$ represents the sign of the price jump that occurs for each unique LOB event. From the LOB events defined in Section 2.1., we know that if the LOB event simulated by the intensity function is in the set $O_u$, the jump size is positive i.e., $\operatorname{sgn}(\Delta V(t))=1$. Similarly, if the LOB event simulated by the intensity function is in the set $O_d$ or $O_n$, $\operatorname{sgn}(\Delta V(t))=-1$ or $\operatorname{sgn}(\Delta V(t))=0$, respectively.

\begin{figure}[ht]
	\centering
	\includegraphics[width=0.99\linewidth]{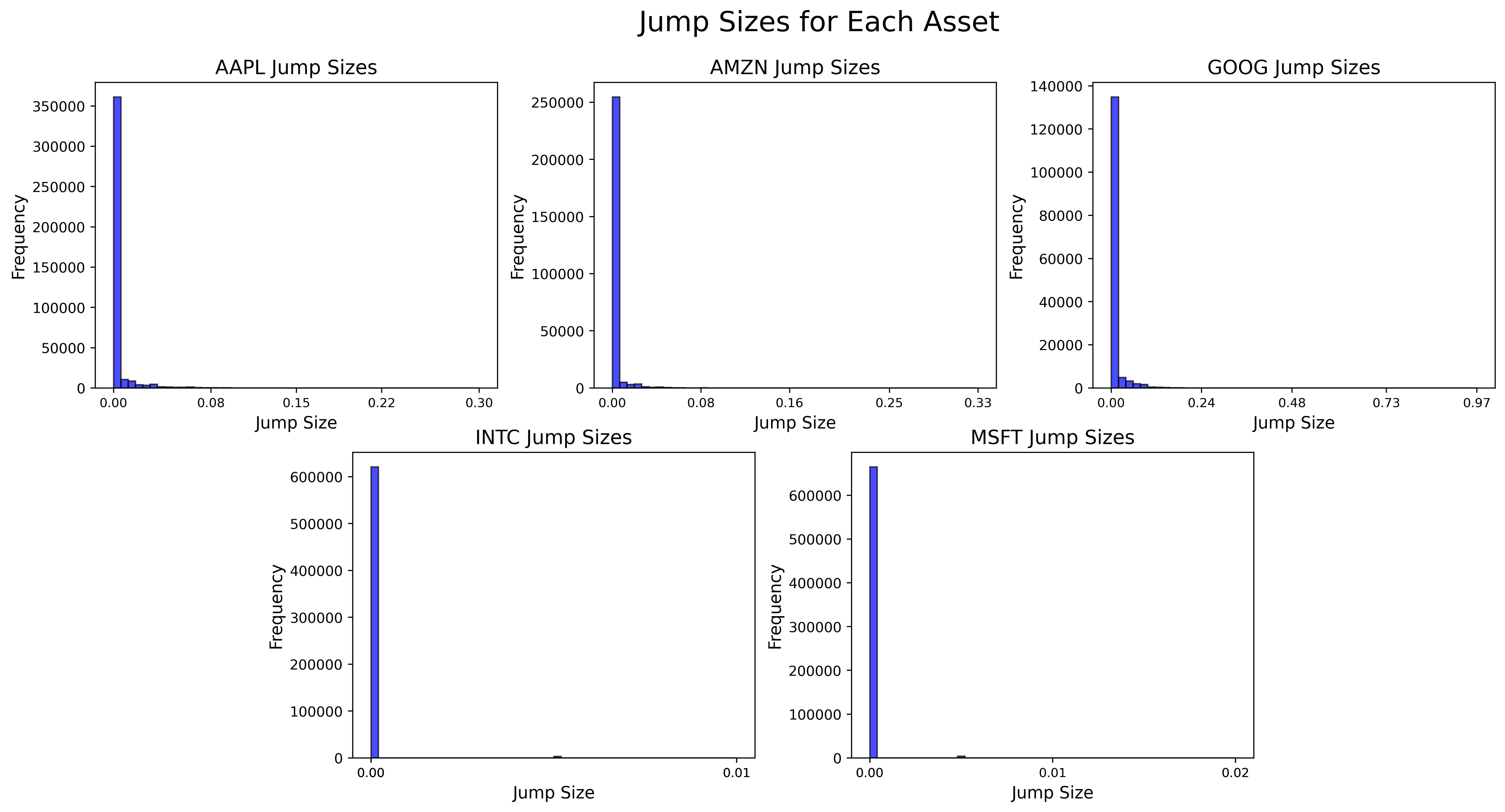}
	\caption{A histogram of jump sizes in each asset studied, showing the extreme heavy right-skewed nature of jump sizes. }
	\label{fig:Jump_sizes}
\end{figure}

Next, we must define the jump size, $\Delta V(t)$,  for each LOB event that has just occurred in order to simulate a midprice process. Recall from Equation (\ref{eq: midprice_process}) that the size of the jumps reflects a mapping from the current state to a price change, where the mapping represents the type of price change. For this particular simulation, we sample our jump sizes, $\Delta V(t)$, from a discrete distribution and this can include $n$ countably finite different jump sizes, where each jump size has a unique probability of occurring, where $n$ would be interpreted as the total number of different jump sizes. Thus, we define the probability of a jump size being a particular size as follows, 
\begin{equation}
P(\Delta V(t) = x) =
\begin{cases}
p_1,  & x = j_1, \\
p_2,  & x = j_2, \\
\vdots & \vdots \\
p_n,  & x = j_n,
\end{cases}
\label{eq:jump_distribution_n}
\end{equation}
where $p_i$ and $j_i$ represent the probability and size of each jump that can occur, respectively. With real LOB data, this can vary significantly depending on the financial asset being modeled. For example, there are many assets where a large portion of the jump moves are 1 tick, and a very small portion of jump sizes have tick size 2, 3, 4,..., etc. Thus, these jump distributions are often heavily skewed, and each simulation process must accurately reflect this for the particular asset being modeled/simulated. See evidence of this in Figure \ref{fig:Jump_sizes} from the data we studied, where one can also see the total number of different jump sizes in Table \ref{tab:jump_sizes_and_stats}. It is important to note that no jump size value can be deemed impossible for a particular asset, but above a certain size is extremely unlikely in most most market regimes; thus, we ignore this case. An agent or practitioner stress-testing a strategy under this type of model should consider scenarios that allow for more extreme jump events, either by using simulated data or employing datasets with higher volatility than the one used in our study. 

\begin{table}[h]
    \centering
    \begin{tabular}{lcccccc}
        \toprule
         & AAPL & AMZN & GOOG & INTC & MSFT \\ 
        \midrule
        Jump sizes ($n$) & 50 & 47  & 91 &  2 & 3   \\  
        \midrule
        Volatility (Real)  & 0.00005 & 0.0001 & 0.00011 & 0.00018 & 0.00016 \\  
        Volatility (Sim) & 0.00005 & 0.0001 & 0.0002 & 0.0001 & 0.0002 \\  
        \midrule
        Absolute Skewness (Real) & 0.0685 & 0.1276 & 0.0267 & 0.1014 & 0.0908  \\  
        Absolute Skewness (Sim)  & 0.1688 & 0.3134 & 1.6717 & 0.0144 & 0.0555  \\  
        \midrule
        Excess Kurtosis (Real)  & 6.6918 & 17.7054 & 35.396 & -1.9814 & -1.9156 \\  
        Excess Kurtosis (Sim) & 6.0606 & 13.3366 & 33.8665 & -1.9896 & -1.9413 \\  
        \midrule
        Hurst Exponent (Real)  & 0.3929 & 0.3428 & 0.3415 & 0.1874 & 0.2846 \\  
        Hurst Exponent (Sim) & 0.5759 & 0.6521 & 0.5707 & 0.3274 & 0.5583 \\  
        \bottomrule
    \end{tabular}
    \caption{Number of different jump sizes in each asset on June 21st, 2012, along with stylized statistics for the log returns in the real and simulated data.}
    \label{tab:jump_sizes_and_stats}
\end{table}

\begin{figure}[H]
	\centering
	\includegraphics[width=0.99\linewidth]{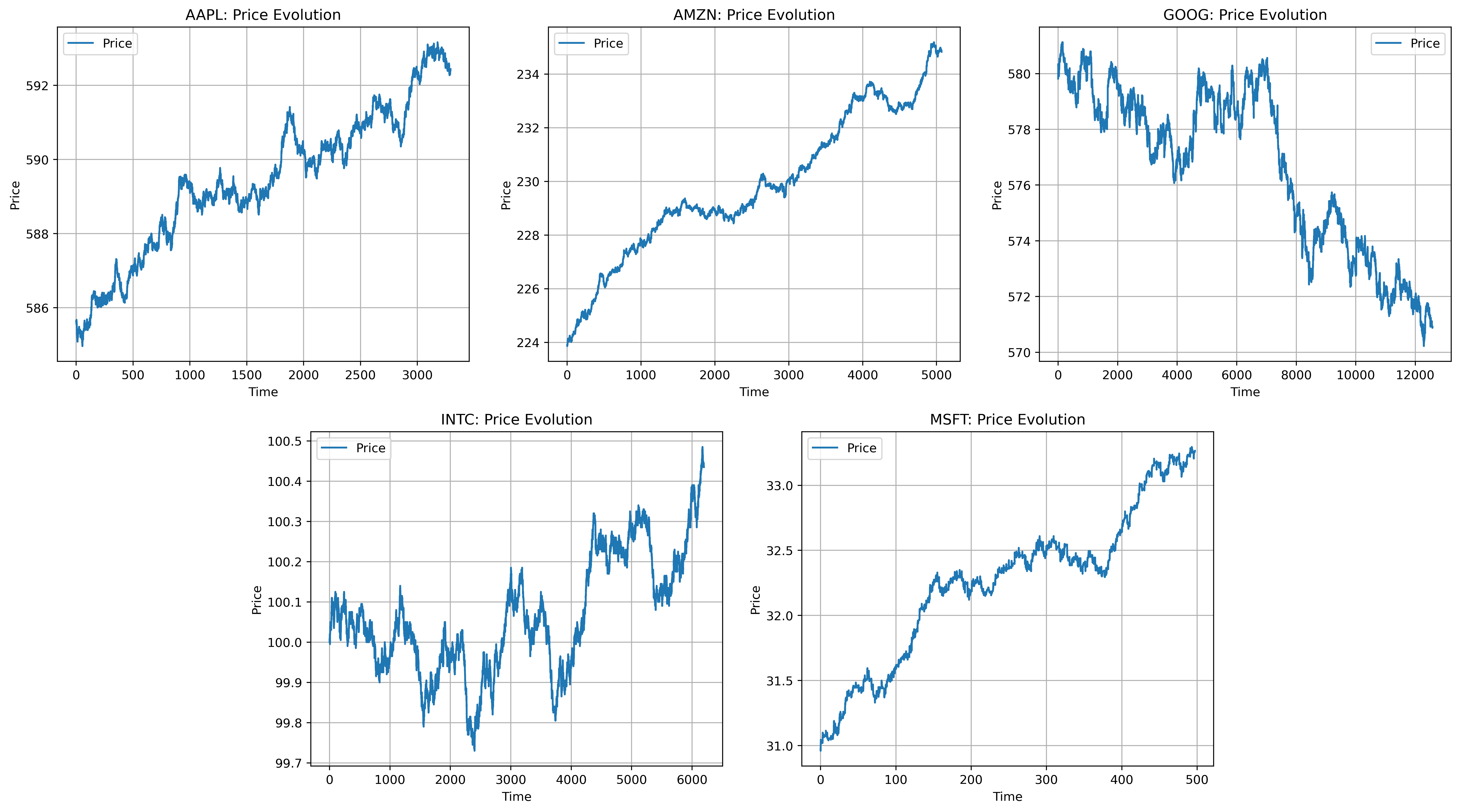}
	\caption{The evolution of the midprice process in each asset based on the previous LOB event simulations.}
	\label{fig:Sample_Price_Path_Sim}
\end{figure}

In Figure \ref{fig:Sample_Price_Path_Sim} we provide a visualization of how each price path evolved in all five assets, and in Table \ref{tab:jump_sizes_and_stats} one can find a comparison of some common high-frequency asset price metrics between this simulated data and the real data. It is pretty clear from the results in Table \ref{tab:jump_sizes_and_stats} that the simulated data was able to capture very similar measures in volatility and excess kurtosis, but there were measurable differences in the absolute skewness and Hurst Exponent values. Thus, we can conclude that our simulated data effectively captures broader characteristics such as overall volatility and the fat-tailed nature of the distribution. However, it performs less accurately in finer details, such as matching the observed price range and distinguishing between trending and mean-reverting market behavior.

\sloppy
\section{Application: Deep Reinforcement Learning Market-Making Problem}
\fussy
In this Section, we study how one can back-test a Market-Making style strategy under the previous LOB simulation process, where we will also compare the results with what would have occurred under the real data. One of the major advantages of using an event-based LOB simulation process is that we can now match trade order fills with times they actually occurred, which is not possible under the commonly used diffusion, pure jump, or jump-diffusion models where these types of LOB events are either not tracked or deemed to be independent of the price process. For example, if an MM has a limit order posted at the best bid, we now know a trade can only take place if a Market Order enters the market, which would be recognizable via the two previously defined Sell Market Order types, $MS^-$ and $MS^0$. Thus, our MM simulation ensures that a trade order fill only occurs when two parties are present, reflecting the fundamental requirement of real market interactions.

To briefly summarize a MM strategy, consider a market participant who posts limit orders at the bid and ask prices. For simplicity, we assume these orders are placed at the best bid and ask prices. The MM is then seen as providing liquidity to the market as this encourages and enables other market participants to trade at current prices. Large MM players often receive fees from the exchanges for doing this, thus this type of activity can generate positive rewards. In saying that, the MM is still exposed to market risk, thus it is in their best interest to figure out when it is optimal to have limit orders posted in the LOB. In terms of optimization, the goal of the MM is to maximize their terminal reward subject to certain constraints that they may have. These types of constraints often include meeting certain risk measures, such as a maximum position size. In this section, we will explore a common optimization-type MM problem under a deep RL framework, where we aim to compute the optimal limit order posting strategy. This optimal strategy will be largely determined by the LOB model in Section 3, as price movements and the timing of the MM's limit order fills are two of the main factors influencing the strategy's overall performance. 

In recent times, deep RL has become one of the more popular optimization approaches to solving these types of Algorithmic and HFT Trading problems. Previously, using Stochastic Optimal Control theory was a common approach in the literature, where popular works include the \cite{bertsimas1998optimal} optimal execution framework, the \cite{bouchard2011optimal} general impulse approach, the \cite{cartea2015algorithmic} textbook which devotes a full chapter to many variations of the MM problem, \cite{gueant2017optimal} by giving a very thorough theoretical overview of the optimal MM problem, and \cite{cartea2018algorithmic} which studies parts of the adverse selection problem in MM. Deep RL gained popularity as advancements in deep learning made it easier to integrate function approximators into the traditional RL framework. This overcame a major limitation of classical RL, which struggled with large state-action spaces commonly found in MM-style problems. See \cite{lalor2024reinforcement} for a more in-depth discussion on the advantages of deep RL versus stochastic optimal control. Examples in the literature of some recent deep RL applications in MM include \cite{gueant2019deep} using deep neural networks under an Actor-Critic method for optimal MM in corporate bonds, \cite{gavsperov2021reinforcement} provides an overview of popular deep RL approaches in optimal MM, \cite{gavsperov2022deep} solves a Hawkes-based optimal MM problem under the Soft Actor-Critic (SAC) approach, \cite{kumar2024deep} formulates an agent-based deep Hawkes model for high-frequency MM and \cite{lalor2024reinforcement} uses the SAC method under non-Markov based price processes.

The rest of this section will proceed as follows. In Section 4.1, we will begin by discussing the common MM setup that we studied, specifically describing the main components of the deep RL framework. This will involve defining the stochastic processes involved, along with how the trading strategy will be simulated. Then, in Section 4.2, we will provide simulation results for the MM strategy under our simulated LOB setup, as discussed in Section 3.2, as well as showing how the strategy would have performed under the real data. To the best of our knowledge, this will be the first extension of a Neural Hawkes event-based LOB model applied to an MM strategy to date. 

\subsection{\secondlevelheading{Market-Making Setup}}
The agent in this setting will be able to post limit orders on the best bid and ask, where their goal will be to maximize their terminal reward subject to an inventory constraint. In a deep RL MM framework, the main variables that must be formulated are the state-action space (to be approximated by a Neural Network) and the reward function. Here, we will follow a similar approach to the deep RL setup in \cite{gavsperov2022deep} and \cite{lalor2024reinforcement}. This type of MM strategy was also studied under the more traditional Stochastic Optimal Control approach in \cite{cartea2015algorithmic} and \cite{cartea2018algorithmic}. First we will define the state space as follows: 
\begin{align}
    S_t = (V_t, Q_t), 
    \label{eq: state_space}
\end{align}
where $V_t$ is the midprice process given in Equation (\ref{eq: midprice_process}) and $Q_t$ will be the inventory process that satisfies the follow equation:
\begin{align}
    Q_t = N_t^--N_t^+,
    \label{eq: inventory_process}
\end{align}
where $N_t$ is a counting process for the limit order fills on the bid ($-$) and the ask ($+$). In our model, each of MM's limit order fills must now coincide with the 4 market order events in Table \ref{tab:LOB_Event_Counts}. Thus, this trade order fill process logic will be extended to include the event-based nature of a trade order fill. 

Trade order fills will be divided into separate non-overlapping counting processes for non-adverse fills and adverse fills, as in \cite{lalor2024market} and \cite{lalor2024reinforcement}, but now following our event-based logic. 
The non-adverse trade order fill process will also attach the same type of non-adverse fill probability, as limit order trade fills are not guaranteed to occur when trades are executed at the agent's limit order price by non-aggressive market orders. This is essentially due to the time-priority nature of LOBs, whereby each limit order is generally sent to the back of a queue of previously submitted limit orders. Thus, only market orders that are greater than or equal to the size of this queue are guaranteed to lead to a limit order fill. 

Subsequently, we first define the trade order fill logic for non-adverse trade order fills as follows: 
\begin{align}
    NFA_t = \sum_{i=1}^NA^+_{t_i}\mathbb{I}_{\{MB^0_{t_i}\neq\emptyset\}}*p,
\end{align}
and
\begin{align}
    NFB_t = \sum_{i=1}^NA^-_{t_i}\mathbb{I}_{\{MS^0_{t_i}\neq\emptyset\}}*p,
\end{align}
where $NFA_t$ and $NFB_t$ represent the counting processes for all non-adverse fills that occur on the best ask and best bid, respectively. Here, $A^+_{t_i}$ and $A^-_{t_i}$ denote trades that would've occurred on the best ask/bid, given the non-aggressive market order events, $MB^0$ and $MS^0$, occurred at time $t_i$. However, the MM agent is not guaranteed to receive these fills so we also attached a non-adverse fill probability, $p$, which can be defined as,
\begin{align}
    p = P(N_{t_i}^{A, \pm}|A_{t_i}^+\mathbb{I}_{\{MB^0_{t_i}\neq\emptyset\}}=0, A_{t_i}^-\mathbb{I}_{\{MS^0_{t_i}\neq\emptyset\}}=0).
\end{align}
This represents the probability of getting filled given that a non-aggressive market order has entered the market. Next, adverse fills, where the agent is guaranteed to have their limit order filled if an aggressive market order enters the market, can be defined as follows,
\begin{align}
    AFA_{t} = \sum_{i=1}^NA_{t_i}^+\mathbb{I}_{\{MB^+_{t_i}\neq\emptyset\}},
\end{align}
and
\begin{align}
    AFB_{t} = \sum_{i=1}^NA_{t_i}^-\mathbb{I}_{\{MS^-_{t_i}\neq\emptyset\}},
\end{align}
where $AFA_t$ and $AFB_t$ represent the counting processes for all adverse fills that occur on the best ask and bid, respectively. The main difference in the trade order fill logic in this work compared to \cite{lalor2024market} and \cite{lalor2024reinforcement} is that now each trade order fill is based on the event-based process of trade orders entering the market, rather than being purely based on simulated price movements from some approximated price process. Intuitively, this new trade order fill logic now states that the agents limit orders will always be filled by aggressive market orders ($MB^+$ and $MS^-$), but only sometimes by non-aggressive market orders ($MB^0$ and $MS^0$), which depends on the non-adverse fill probability. 

Next, define the action space as follows,
\begin{align}
A_t = 
\begin{cases} 
\{0, 1\}, & \text{if } Q_t = -q, \\
\{-1, 0, 1\}, & \text{if } -q < Q_t < q, \\
\{-1, 0\}, & \text{if } Q_t = q. \\
\end{cases}
\label{eq: action_space}
\end{align}
where we can see that the MM agent can choose whether or not to post limit orders at the best bid/ask, subject to an inventory constraint $q$. Thus, the agent cannot have a long or short position larger than the absolute value of $q$.   

Next, the reward function must be defined, in order to measure the performance of the deep RL MM strategy. Here we will use the commonly defined reward/value function used in the MM literature as follows,
\begin{align}
    \mathbb{E}_\pi \left[W_T-\psi\int_0^T|Q_t|dt\right],
    \label{eq: reward_func}
\end{align}
where $W_t$ represents the total wealth which can be defined as $W_t = Q_t V_t + C_t$, where $C_t$ represents the cash process. Here, $\psi$ is called the inventory penalty coefficient, which penalizes inventories greater than zero. A MM agent generally prefers to keep their positions in the market low or large only for a very short period, thus, this will encourage the model to close out any large positions as quickly as possible. 

Lastly, we will briefly describe the Neural Network architecture used in approximating the State-Action space. As previously stated, we use the Soft Actor-Critic (SAC) approach, which is an off-policy deep RL algorithm that maximizes the expected reward while encouraging exploration through entropy regularization, using two Q-networks, a stochastic policy, and a automatic parameter adjustment for balancing exploitation and exploration. We used the Stable Baselines3 python package to streamline this, which was recently developed by \cite{stable-baselines3}. See \cite{haarnoja2018soft} for the seminal paper on this algorithm and \cite{lalor2024reinforcement} for a more in-depth introduction on how to apply this in a MM setup as we use the same approach. 

\subsection{\secondlevelheading{Asset Specific Results - Simulated vs Real Data}}

In this Section, we will discuss the results that pertain to an asset-specific approach, where each set of results is based on the LOB dynamics learned from the data of each specific asset. We conducted a comparative analysis between the output of the simulated midprice process (based on the Neural Hawkes process framework described in Section 3) and the real \cite{LOBSTER} data. In both cases, the deep RL market-making model was trained separately on each of the five assets. From this analysis, we are able to assess the cumulative performance results on unseen data for each particular asset and we can analyze how the major market order event types impacted the MM agent's performance.

First, see Table \ref{tab:Deep_RL_Params}, where we show the parameter values used in our Deep RL MM model. To start with, we set the maximum inventory to $5$ units to prevent the agent's position from becoming too large. Smaller positions are further encouraged by the inventory penalty parameter, where we set $\psi = 0.001$. Each trade is of size $1$, represented by $dQ = \pm 1$. Then we set the non-adverse fill probability to $0.2$, which is in line with the empirical results in \cite{lalor2024market}, where they computed this probability based on trade order fill data on some of the most liquid futures contracts listed on the Chicago Mercantile Exchange. Lastly, the training and testing sample datasets, denoted $X_{train}$ and $X_{test}$, were performed on the first $5,000$ and the subsequent $2,500$ samples of LOB event and price data, respectively, for both the simulated and real datasets.  

\begin{table}[ht!]
	\begin{center}
		\begin{tabular}{ |p{2.5cm}|p{2.5cm}|p{2.5cm}|p{2.5cm}|  }
 		\hline
 		\multicolumn{4}{|c|}{\textbf{Parameters}} \\
 		\hline
 		\textbf{Parameter} & \textbf{Value} 		&\textbf{Parameter} & \textbf{Value}\\
 		\hline
      	$q$ & $5$ & dQ & $\pm 1$ \\
            $p$ & $0.2$ & $\psi$ & $0.001$\\
            $X_{train}$ & $5,000$ & $X_{test}$ & $2,500$\\
 		\hline
		\end{tabular}
	\captionsetup{font=small}
	\caption{Deep RL optimal MM simulation parameters.}
	\label{tab:Deep_RL_Params}
	\end{center}
\end{table}

\begin{figure}[H]
	\centering
	\includegraphics[width=0.99\linewidth]{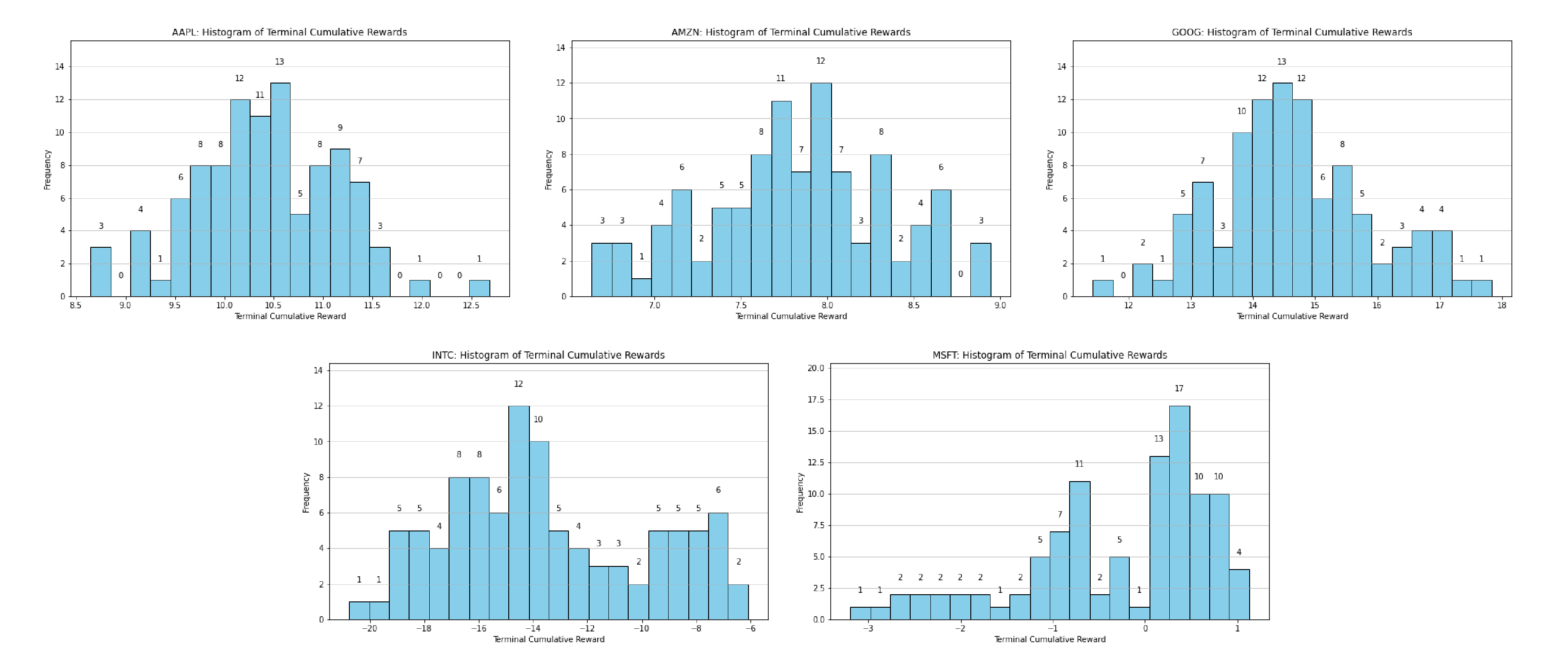}
	\caption{A histogram of the terminal cumulative reward over each testing episode from the \textbf{simulated} data.}
	\label{fig:Hist_Sim}
\end{figure}

\begin{figure}[H]
	\centering
	\includegraphics[width=0.99\linewidth]{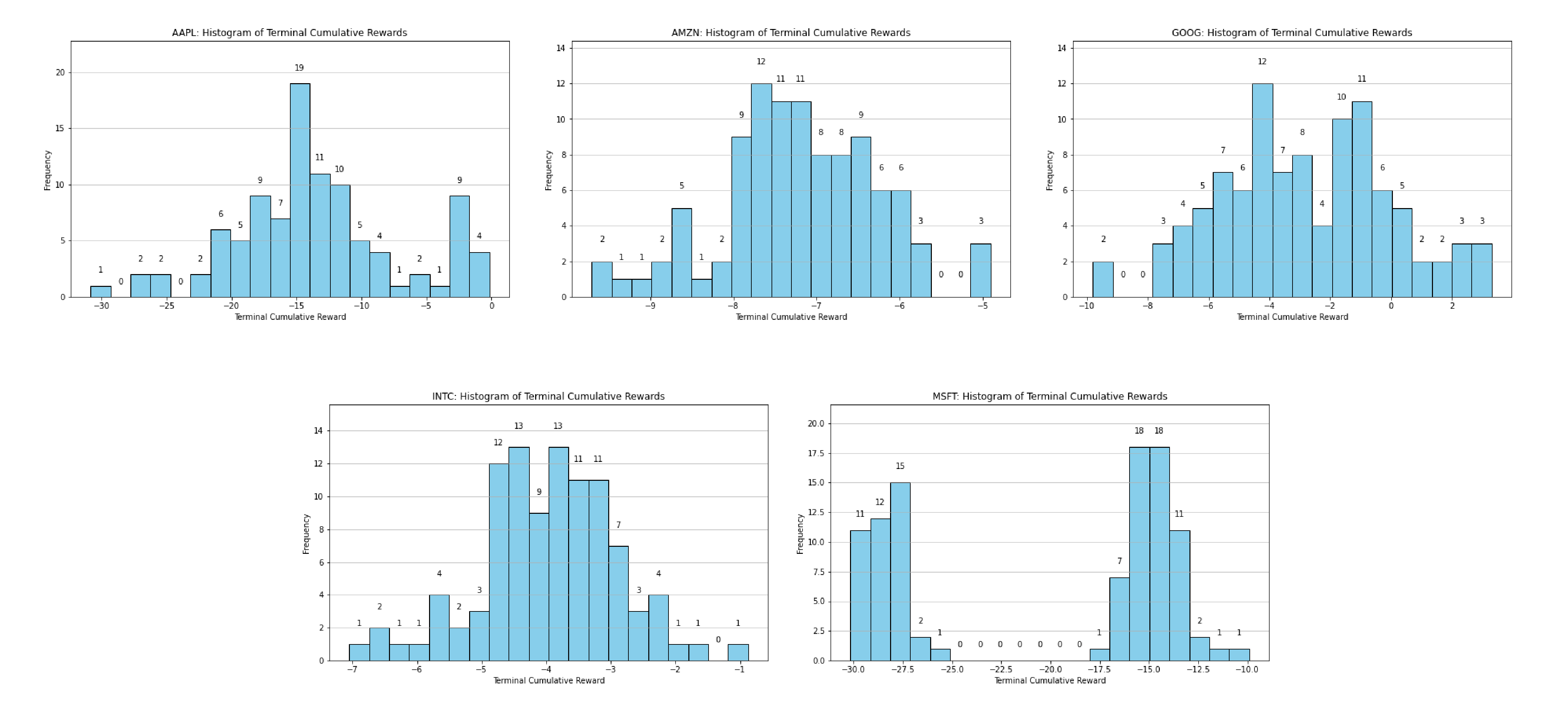}
	\caption{A histogram of the terminal cumulative reward over each testing episode from the \textbf{real} data.}
	\label{fig:Hist_Real}
\end{figure}

We begin by presenting the key cumulative performance results from this analysis, which in RL is often done by analyzing the output of the reward function at the terminal time $T$. Figures \ref{fig:Hist_Sim} and \ref{fig:Hist_Real} illustrate the cumulative terminal rewards in all 100 testing episodes, for both the simulated and real data. Here, it is clear that the cumulative rewards are mostly negative for both sets of data, with some slight differences in how these terminal value outcomes were distributed. This still represents an improvement over the initial results observed during the training period, similar to the findings in \cite{lalor2024reinforcement}, where the MM agent successfully learned to achieve a positive or a less negative reward. However, it is pretty clear that a simple strategy, such as one purely focused on finding optimal times to post limit orders on the best bid/ask, is unlikely to be profitable on its own. The secretive nature of high-performing MM strategies makes it tough to improve on more traditional MM setups, but the main point here is to show how our event-based LOB model can be applied in a real-world setting. 

\begin{figure}[H]
	\centering
	\includegraphics[width=0.99\linewidth]{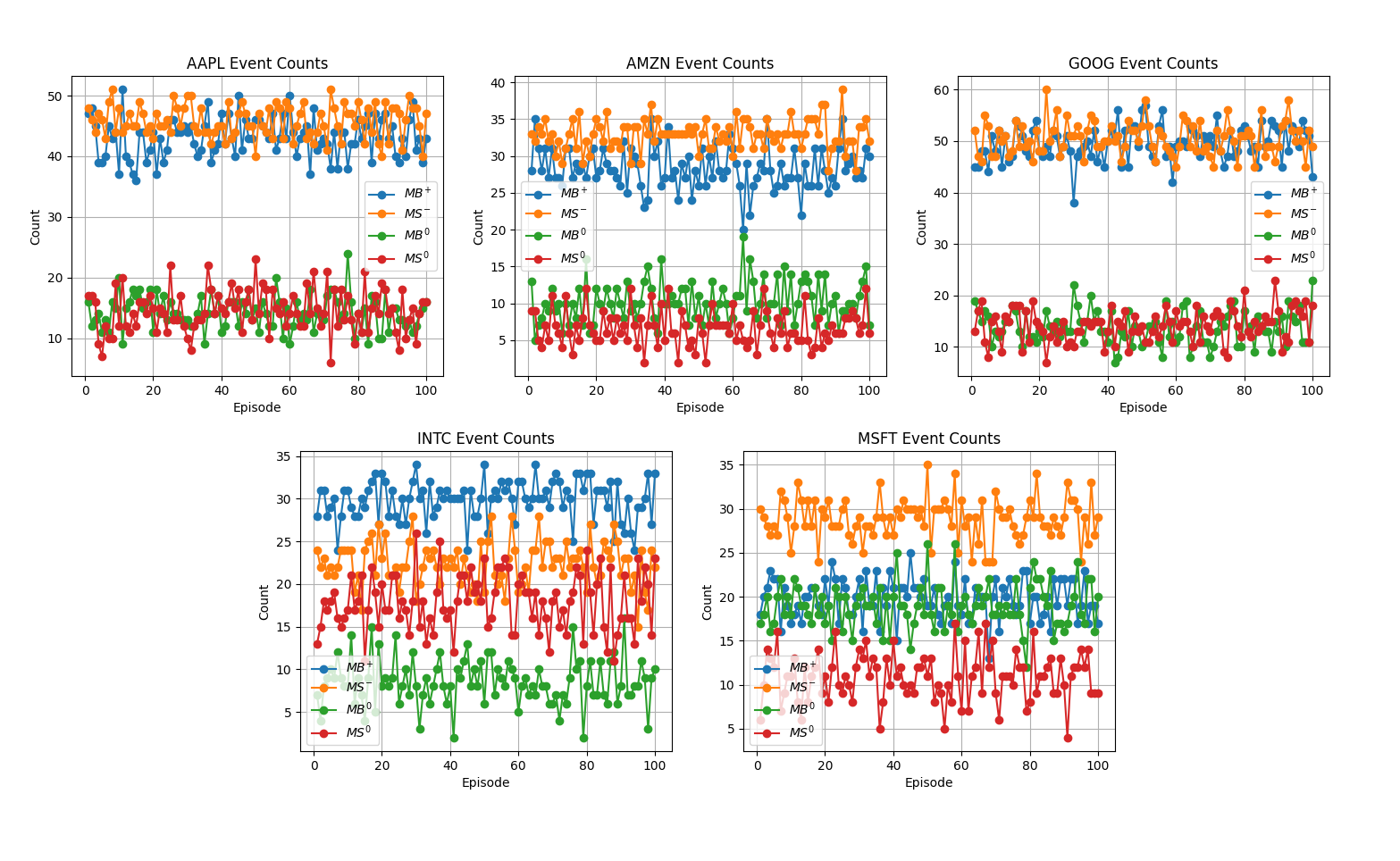}
	\caption{The number of times each Market Order type resulted in a trade order fill for the MM agent in each testing episode in the \textbf{simulated} data.}
	\label{fig:Event_Counts_RL_Sim}
\end{figure}

\begin{figure}[H]
	\centering
	\includegraphics[width=0.99\linewidth]{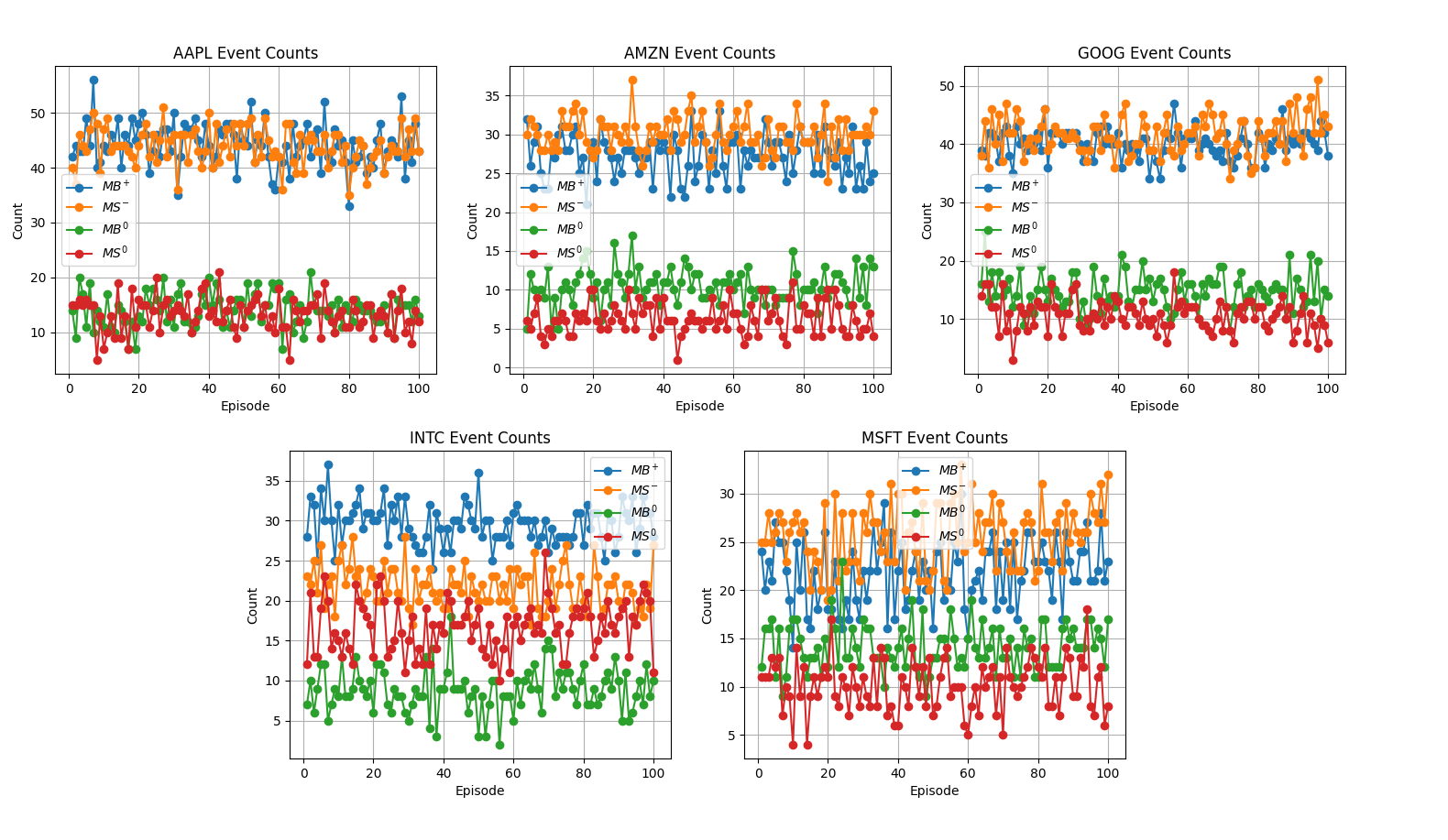}
	\caption{The number of times each Market Order type resulted in a trade order fill for the MM agent in each testing episode in the \textbf{real} data.}
	\label{fig:Event_Counts_RL_Real}
\end{figure}

As noted previously, the performance of an MM strategy is often heavily determined by how and when the MM agent receives limit order trade fills and, in particular, how these trade order fills affect the general market exposure of the MM agent. Given our event-based framework, we can now analyze which of the previously defined market order types, as shown in Table \ref{tab:LOB_Event_Counts}, matched with the MM's agent's posted limit orders. More specifically, Figures \ref{fig:Event_Counts_RL_Sim} and \ref{fig:Event_Counts_RL_Real} illustrate the frequency with which each type of market order filled the agent's posted limit orders across all testing episodes, in both simulated and real data, respectively. It is clear that the distribution of the MM agents limit order trade fills is much more likely to come from aggressive market orders rather than non-aggressive market orders, and this is very similar in both the simulated and real data. See also Table \ref{tab:Adv_NonAdv_Ratio}, where one can also see that the ratio between adverse and non-adverse trade order fills were very similar in each asset, as well as being significantly weighted toward adverse fills. This essentially means that after the MM agent receives a new trade order fill, the resulting position will be out of the money at the next time step more often than not. In reality, it would be in the MM agent's best interest to avoid these adverse limit order fills. However, as numerous studies have shown, including \cite{lalor2024market}, effectively mitigating them remains a significant challenge.

\begin{table}[h]
    \centering
    \begin{tabular}{lcccccc}
        \toprule
         & AAPL & AMZN & GOOG & INTC & MSFT \\   
        \midrule
        Ratio (Real) & 3.2093 & 3.4203 & 3.2342 & 2.0004 & 1.9510  \\  
        Ratio (Sim) & 3.0871 & 3.4626 & 3.4151 & 1.9708 & 1.6075 \\  
         
        \bottomrule
    \end{tabular}
    \caption{The ratio of the MMs adverse to non-adverse trade order fills for each asset across all the Neural Hawkes simulation and real data testing episodes.}
    \label{tab:Adv_NonAdv_Ratio}
\end{table}

\section{Conclusions and Future Recommendations}

In this research, we developed an event-based Neural HP for simulating asset price process at a high-frequency level. Granular LOB information is essential for HFT strategies like for a MM, and we believe our model takes strides in improving analysis in this area. More specifically, we developed an event-based LOB model that takes into account 12 of the main events that appear in the LOB, where each event's intensity was modeled via a nonlinear MVHP. This nonlinear MVHP model was then approximated via the Neural HP, which helps to overcome many of the limitations present in the traditional HP models. Our empirical results show that many of the broad dynamics seen in the simulated data are in line with the real data, as seen by the volatility and excess kurtosis measures, while the model still struggles with some of the finer details, evidenced by the absolute skewness and Hurst exponent measures. However, the resulting simulated midprice process was then applied with a deep RL MM framework, where very similar results were achieved under the simulated and real data. In this setting, the MMs trade order fills can now be specifically calculated from the LOB event activity observed, which is more in line with how a high-frequency MM assesses market conditions in the real world. 

In terms of future research recommendations, we recommend digging deeper to improve the accuracy of the predicted simulated midprice processes, so that it can align even more closely with the real LOB data. Although our Neural Hawkes setup yielded similar results for the simple MM strategy across both simulated and real data, enhancing the model to better distinguish between aggressive and non-aggressive orders would lead to a more accurate representation of the high-frequency dynamics observed in LOB data. A practitioner may also want to get even more specific with the set of LOB events, by extending the LOB model to include more than the 12 events currently included. This should be achievable if the LOB event's impact on the midprice process and the dynamics of its intensity function can be mathematically formulated in a reasonably simple way. 

\section*{Acknowledgments}
The authors thank MITACS and NSERC for research funding.

\section*{Declarations of Interest}
The authors declare that they have no conflicts of interest.  

\bibliographystyle{chicago}
\bibliography{Paper1}

\begin{thebibliography}{}

\bibitem[Abergel and Jedidi, 2015]{abergel2015long}
Abergel, F. and Jedidi, A. (2015).
\newblock Long-time behavior of a hawkes process--based limit order book.
\newblock {\em SIAM Journal on Financial Mathematics}, 6(1):1026--1043.

\bibitem[Bacry et~al., 2015]{bacry2015hawkes}
Bacry, E., Mastromatteo, I., and Muzy, J.-F. (2015).
\newblock Hawkes processes in finance.
\newblock {\em Market Microstructure and Liquidity}, 1(01):1550005.

\bibitem[Bertsimas and Lo, 1998]{bertsimas1998optimal}
Bertsimas, D. and Lo, A.~W. (1998).
\newblock Optimal control of execution costs.
\newblock {\em Journal of financial markets}, 1(1):1--50.

\bibitem[Black and Scholes, 1973]{black1973pricing}
Black, F. and Scholes, M. (1973).
\newblock The pricing of options and corporate liabilities.
\newblock {\em Journal of political economy}, 81(3):637--654.

\bibitem[Bouchard et~al., 2011]{bouchard2011optimal}
Bouchard, B., Dang, N.-M., and Lehalle, C.-A. (2011).
\newblock Optimal control of trading algorithms: a general impulse control approach.
\newblock {\em SIAM Journal on financial mathematics}, 2(1):404--438.

\bibitem[Cartea et~al., 2018a]{cartea2018enhancing}
Cartea, A., Donnelly, R., and Jaimungal, S. (2018a).
\newblock Enhancing trading strategies with order book signals.
\newblock {\em Applied Mathematical Finance}, 25(1):1--35.

\bibitem[Cartea et~al., 2015]{cartea2015algorithmic}
Cartea, {\'A}., Jaimungal, S., and Penalva, J. (2015).
\newblock {\em Algorithmic and high-frequency trading}.
\newblock Cambridge University Press.

\bibitem[Cartea et~al., 2018b]{cartea2018algorithmic}
Cartea, A., Jaimungal, S., and Ricci, J. (2018b).
\newblock Algorithmic trading, stochastic control, and mutually exciting processes.
\newblock {\em SIAM review}, 60(3):673--703.

\bibitem[DeLise, 2024]{delise2024negative}
DeLise, T. (2024).
\newblock The negative drift of a limit order fill.
\newblock {\em arXiv preprint arXiv:2407.16527}.

\bibitem[Ga{\v{s}}perov et~al., 2021]{gavsperov2021reinforcement}
Ga{\v{s}}perov, B., Begu{\v{s}}i{\'c}, S., Posedel~{\v{S}}imovi{\'c}, P., and Kostanj{\v{c}}ar, Z. (2021).
\newblock Reinforcement learning approaches to optimal market making.
\newblock {\em Mathematics}, 9(21):2689.

\bibitem[Ga{\v{s}}perov and Kostanj{\v{c}}ar, 2022]{gavsperov2022deep}
Ga{\v{s}}perov, B. and Kostanj{\v{c}}ar, Z. (2022).
\newblock Deep reinforcement learning for market making under a {Hawkes} process-based limit order book model.
\newblock {\em IEEE control systems letters}, 6:2485--2490.

\bibitem[Gould et~al., 2013]{gould2013limit}
Gould, M.~D., Porter, M.~A., Williams, S., McDonald, M., Fenn, D.~J., and Howison, S.~D. (2013).
\newblock Limit order books.
\newblock {\em Quantitative Finance}, 13(11):1709--1742.

\bibitem[Gu{\'e}ant, 2017]{gueant2017optimal}
Gu{\'e}ant, O. (2017).
\newblock Optimal market making.
\newblock {\em Applied Mathematical Finance}, 24(2):112--154.

\bibitem[Gu{\'e}ant and Manziuk, 2019]{gueant2019deep}
Gu{\'e}ant, O. and Manziuk, I. (2019).
\newblock Deep reinforcement learning for market making in corporate bonds: beating the curse of dimensionality.
\newblock {\em Applied Mathematical Finance}, 26(5):387--452.

\bibitem[Haarnoja et~al., 2018]{haarnoja2018soft}
Haarnoja, T., Zhou, A., Abbeel, P., and Levine, S. (2018).
\newblock Soft actor-critic: Off-policy maximum entropy deep reinforcement learning with a stochastic actor.
\newblock In {\em International conference on machine learning}, pages 1861--1870. PMLR.

\bibitem[Hochreiter, 1997]{hochreiter1997long}
Hochreiter, S. (1997).
\newblock Long short-term memory.
\newblock {\em Neural Computation MIT-Press}.

\bibitem[Jain et~al., 2024]{jain2024limit}
Jain, K., Firoozye, N., Kochems, J., and Treleaven, P. (2024).
\newblock Limit order book simulations: A review.
\newblock {\em arXiv preprint arXiv:2402.17359}.

\bibitem[Kloeden et~al., 1992]{kloeden1992stochastic}
Kloeden, P.~E., Platen, E., Kloeden, P.~E., and Platen, E. (1992).
\newblock {\em Stochastic differential equations}.
\newblock Springer.

\bibitem[Kumar, 2024]{kumar2024deep}
Kumar, P. (2024).
\newblock Deep hawkes process for high-frequency market making.
\newblock {\em Journal of Banking and Financial Technology}, pages 1--18.

\bibitem[Lalor and Swishchuk, 2024a]{lalor2024algorithmic}
Lalor, L. and Swishchuk, A. (2024a).
\newblock Algorithmic and high-frequency trading problems for semi-{Markov} and {Hawkes} jump-diffusion models.
\newblock {\em arXiv preprint arXiv:2409.12776}.

\bibitem[Lalor and Swishchuk, 2024b]{lalor2024market}
Lalor, L. and Swishchuk, A. (2024b).
\newblock Market simulation under adverse selection.
\newblock {\em arXiv preprint arXiv:2409.12721}.

\bibitem[Lalor and Swishchuk, 2024c]{lalor2024reinforcement}
Lalor, L. and Swishchuk, A. (2024c).
\newblock Reinforcement learning in non-markov market-making.
\newblock {\em arXiv preprint arXiv:2410.14504}.

\bibitem[Law and Viens, 2019]{law2019market}
Law, B. and Viens, F. (2019).
\newblock Market making under a weakly consistent limit order book model.
\newblock {\em High Frequency}, 2(3-4):215--238.

\bibitem[Lewis and Shedler, 1979]{lewis1979simulation}
Lewis, P.~W. and Shedler, G.~S. (1979).
\newblock Simulation of nonhomogeneous poisson processes by thinning.
\newblock {\em Naval research logistics quarterly}, 26(3):403--413.

\bibitem[Liniger, 2009]{liniger2009multivariate}
Liniger, T. (2009).
\newblock {\em Multivariate hawkes processes}.
\newblock PhD thesis, ETH Zurich.

\bibitem[{LOBSTER}, 2025]{LOBSTER}
{LOBSTER} (2025).
\newblock {LOBSTER: Limit Order Book Reconstruction and Visualization}.
\newblock Accessed: 2025-01-30.

\bibitem[Lu and Abergel, 2018]{lu2018high}
Lu, X. and Abergel, F. (2018).
\newblock High-dimensional hawkes processes for limit order books: modelling, empirical analysis and numerical calibration.
\newblock {\em Quantitative Finance}, 18(2):249--264.

\bibitem[Mei and Eisner, 2017]{mei2017neural}
Mei, H. and Eisner, J.~M. (2017).
\newblock The neural hawkes process: A neurally self-modulating multivariate point process.
\newblock {\em Advances in neural information processing systems}, 30.

\bibitem[Merton, 1976]{merton1976option}
Merton, R.~C. (1976).
\newblock Option pricing when underlying stock returns are discontinuous.
\newblock {\em Journal of financial economics}, 3(1-2):125--144.

\bibitem[Raffin et~al., 2020]{stable-baselines3}
Raffin, A., Hill, A., Gleave, A., Kanervisto, A., Ernestus, M., and Dormann, N. (2020).
\newblock Stable baselines3.
\newblock GitHub repository.

\bibitem[Roldan~Contreras and Swishchuk, 2022]{roldan2022optimal}
Roldan~Contreras, A. and Swishchuk, A. (2022).
\newblock Optimal liquidation, acquisition and market making problems in {HFT} under {Hawkes} models for {LOB}.
\newblock {\em Risks}, 10(8):160.

\bibitem[Shi and Cartlidge, 2022]{shi2022state}
Shi, Z. and Cartlidge, J. (2022).
\newblock State dependent parallel neural hawkes process for limit order book event stream prediction and simulation.
\newblock In {\em Proceedings of the 28th ACM SIGKDD Conference on Knowledge Discovery and Data Mining}, pages 1607--1615.

\bibitem[Swishchuk and Huffman, 2020]{swishchuk2020general}
Swishchuk, A. and Huffman, A. (2020).
\newblock General compound {Hawkes} processes in limit order books.
\newblock {\em Risks}, 8(1):28.

\bibitem[Swishchuk et~al., 2019]{swishchuk2019compound}
Swishchuk, A., Remillard, B., Elliott, R., and Chavez-Casillas, J. (2019).
\newblock Compound {Hawkes} processes in limit order books.
\newblock In {\em Financial Mathematics, Volatility and Covariance Modelling}, pages 191--214. Routledge.

\end{thebibliography}

\appendix

\end{document}